\documentclass[11pt,a4paper]{article}
\pdfoutput=1
\usepackage{graphicx}
\usepackage{jheppub}
\usepackage{amsfonts,amssymb}
\usepackage{amsmath}
\usepackage{epsfig}
\usepackage{bm}

\def\vac{{\rm vac}}
\def\k{{\bm k}}
\def\x{{\bm x}}
\def\bzero{{\bm 0}}

\begin{document}

\title{Absence of a local rest frame in far from equilibrium quantum matter}

\author{Peter Arnold}
\affiliation{Department of Physics, University of Virginia, 382 McCormick Rd, Charlottesville, VA 22904-4714, USA}
\author{Paul Romatschke}
\affiliation{Department of Physics, 390 UCB, University of Colorado, Boulder, CO 80309-0390, USA}
\author{Wilke van der Schee}
\affiliation{Institute for Theoretical Physics and Institute for Subatomic Physics,
Utrecht University, Leuvenlaan 4, 3584 CE Utrecht, The Netherlands}

\date{\today}

\abstract{
In a collision of strongly coupled quantum matter we find that the dynamics of the collision produces regions where a local rest frame cannot be defined because the energy-momentum tensor does not have a real time-like eigenvector. This effect is purely quantum mechanical, since for classical systems, a local rest frame can always be defined. We study the relation with the null and weak energy condition, which are violated in even larger regions, and compare with previously known examples. While no pathologies or instabilities arise, it is interesting that regions without a rest frame are possibly present in heavy ion collisions.
}

\maketitle

\section{Introduction}
\noindent
In condensed matter physics, plasma physics, and nuclear and particle physics, it is often taken for granted that while systems can be out of equilibrium, a local rest frame for matter can always be defined. After all, this seemingly innocuous assumption just states that, at least locally, there should be a frame with respect to which matter is not moving. This is particularly natural in cases of gases or fluids near equilibrium, for which the energy-momentum tensor can be written explicitly using dissipative hydrodynamics. In this work, we report on explicit examples of relativistic quantum systems far from equilibrium, for which one cannot even define a local rest frame everywhere: the energy-momentum tensor is non-diagonalizable by local Lorentz transformations.

In the context of general relativity, energy-momentum tensors that do not have a time-like eigenvector (and hence do not possess a local rest frame) are known to violate the weak and null energy conditions \cite{Hawking:1973uf}. Energy conditions are important for singularity  theorems. As pointed out e.g. by Klinkhammer \cite{Klinkhammer:1991ki}, in the case of weak energy condition violation, singularity theorems may be repaired by replacing the weak energy condition with an averaged weak energy condition \cite{Roman:1986tp, Fewster:2002ne} along particle geodesics. Nevertheless, concrete examples of systems that violate the weak energy condition are few and create considerable interest. For instance, Hawking and Ellis \cite{Hawking:1973uf} refer to non-diagonalizable energy-momentum tensors as `type IV' and note that ``there are no observed fields with energy-momentum tensors of this form''. Roman \cite{Roman:1986tp} gives examples of non-diagonalizable energy-momentum tensors, but they either are not in flat spacetime or they involve unusual boundary conditions (see also \cite{Ford:1990ae}). More recently, Dubovsky {\it et al.} \cite{Dubovsky:2005xd} find systems with non-diagonalizable energy-momentum tensors by studying derivatively coupled scalar field theories with coordinate dependent condensates which also generally contain a superluminally propagating mode.

In this work, we will first adapt some constructions from the literature on energy conditions to give a simple example from free field theory in flat spacetime that will have no local rest frame for the energy-momentum tensor in certain regions. We will then present new examples in the case of a strongly-interacting theory:
${\cal N}=4$ Super-Yang Mills theory at infinite coupling strength and infinite number of colors, studied numerically via gauge/gravity duality  \cite{Chesler:2010bi, Casalderrey-Solana:2013aba,vanderSchee:2013pia, Chesler:2013lia}. The latter examples are the result of a dynamical collision, calculated for a strongly interacting quantum field theory with an explicitly known Lagrangian in flat spacetime with trivial boundary conditions. They are furthermore constructed to resemble heavy ion collisions, and may as such be relevant for physics at RHIC and LHC \cite{Janik:2005zt,Albacete:2008vs,Gubser:2008pc,Grumiller:2008va,Aref'eva:2009wz}.

\section{Local rest frame and energy conditions}
\noindent
In a system consisting of gas of classical, on-shell particles with mass $m$ having a non-negative particle distribution function $f$, the energy-momentum tensor is given by\footnote{This article uses a mostly plus metric tensor $\eta_{\mu\nu}$, Greek letters for spacetime indices and Roman letters for space indices.}
\begin{equation}
T^{\mu\nu}(t,{\bf x})=\int \frac{d^3p}{(2\pi)^3} \frac{p^\mu p^\nu}{p^0} f(t,{\bf x},{\bf p})\,,\quad p^0=\sqrt{{\bf p}^2+m^2}\,.
\end{equation}
Such a system has a local rest frame if there exists a frame with no momentum flow, i.e.\ $T_{0i}=0$. The boost velocity $u^\mu$ required to go back to the lab frame will then provide a time-like eigenvector of $T^{\mu}_{\;\;\nu}$, for which the corresponding eigenvalue $\epsilon$ equals the energy density in the local rest frame: 
\begin{equation}
\label{eigenval}
T^{\mu}_{\;\;\nu} u^\nu = - \epsilon u^\mu \,,
\end{equation}
where $u^\mu$ is normalized as $u^\mu u_\mu = -1$.
The space components of $u^\mu$, or more specifically $v^i\equiv u^i/u^0$, may be interpreted as the velocity with which the matter is flowing with respect to the `laboratory' or global rest frame. Note that for systems that are close to thermal equilibrium, this velocity would correspond to the local fluid velocity, but, as long as there is a local rest frame, it is a well-defined quantity even far away from thermal equilibrium where fluid dynamics does not apply.

For simplicity, let us first consider the system to have enough symmetry in the transverse ($xy$) plane that flow must be in the $z$ direction, i.e.\
$T^{\perp0} = T^{\perp z}=0$ for $\perp = (x,y)$.
In this case, $u^\mu$ has only one independent component and the eigenvalue equation (\ref{eigenval}) can be solved analytically. One finds that a time-like eigenvector and hence a local rest frame (LRF) exists if
\begin{equation}
\label{condition}
\tag{LRF}
|T^{00}+T^{zz}|>2 |T^{0z}|\,.
\end{equation}

For a comparison to other, analogous energy conditions,
the discussion here will be simplified
if we additionally assume that transverse pressures are non-negative
(i.e.\ $T^{\perp\perp}$ has no negative eigenvalues).
Then we find that all time-like observers observe a non-negative energy density (weak energy condition, WEC) if
\begin{equation}
\label{WECcondition}
T^{00}+2 \beta T^{0z}+ \beta^2 T^{zz}\geq 0, \quad \forall\beta\in(-1,1),
\end{equation}
which is equivalent to
\begin{align}
T^{00}+T^{zz} \geq 2 |T^{0z}|; \text{ and }\hspace{1.3cm} \nonumber \\
T^{00}-(T^{0z})^2/T^{zz}\geq 0\,\text{ when } |T^{0z}|\leq|T^{zz}| .
\tag{WEC}
\label{WECcondition2}
\end{align}
Note that $T^{00}\geq 0$ is implied by (\ref{WECcondition}) and
so also follows from (WEC).

A slightly different condition that is sometimes discussed is the null
energy condition (NEC), which is the condition that
$T_{\mu\nu} u^\mu u^\nu \geq 0$ for all null $u^\mu$.
If the transverse pressure is positive we can take ${\bm u}$ in the $z$ direction and therefore take
$\beta=\pm 1$ in eqn.\ (\ref{WECcondition}) to find
\begin{equation}
\label{NECcondition}
T^{00}+T^{zz} \geq 2 |T^{0z}|.
\tag{NEC}
\end{equation}
Interestingly, we find that the NEC is equivalent to the existence of a rest frame (LRF) if $T^{00}+T^{zz}\geq 0$. If $|T^{0z}|\geq|T^{zz}|$ this is also equivalent to the WEC. In section \ref{example} we will nevertheless find an example of a region with an LRF with  $T^{00}+T^{zz}< 0$ which violates the NEC. In that case the LRF condition is more difficult to violate than the NEC, and the rest frame velocity is in opposite direction of the flux.

For a gas of classical particles with non-negative $f$, both $T^{00}$ and $T^{zz}$ are manifestly positive, so the LRF and NEC inequalities become
$$
\int \frac{d^3p}{(2\pi)^3} \frac{(p^0-p^z)^2}{p^0} f(t,{\bf x},{\bf p}) > 0\,,
$$
which is always fulfilled. Similarly the WEC is always satisfied.  Moreover, a classical scalar field $\phi(t,z)$ with a Lagrangian density ${\cal L}=\frac{1}{2}\partial_\mu \phi \partial^\mu \phi-U(\phi)$ and arbitrary potential $U(\phi)$ can be shown to have $T^{00}+T^{zz}=(\partial_0 \phi)^2+(\partial_z\phi)^2$ and $T^{0z}=\partial^0 \phi \partial^z \phi$, thus always possessing an LRF and satisfying the NEC. If the potential is bounded from below and normalized so that $U(\phi)=0$ at its minimum, also the WEC is satisfied. So for both examples, a gas of classical particles and a system of classical fields, a local rest frame may always be defined, and the weak energy condition is never violated.

\section{An example in free field theory \label{freescalar}} 
\noindent
Before discussing the case of strongly coupled matter, we first give a simple
example in free field theory that violates the LRF condition in some
places and so
has spacetime regions where there is no local rest frame for
$T^{\mu\nu}$.  There is a long history of studying violation of the other energy
conditions in free field theory, and we will adapt a construction from
Davies and Fulling \cite{FullingDavies:1976, Davies:1977yv}
in order to provide an example that violates the LRF condition
as well.

\begin{figure*}[t!]
  \centering
  \includegraphics[height=0.52\linewidth]{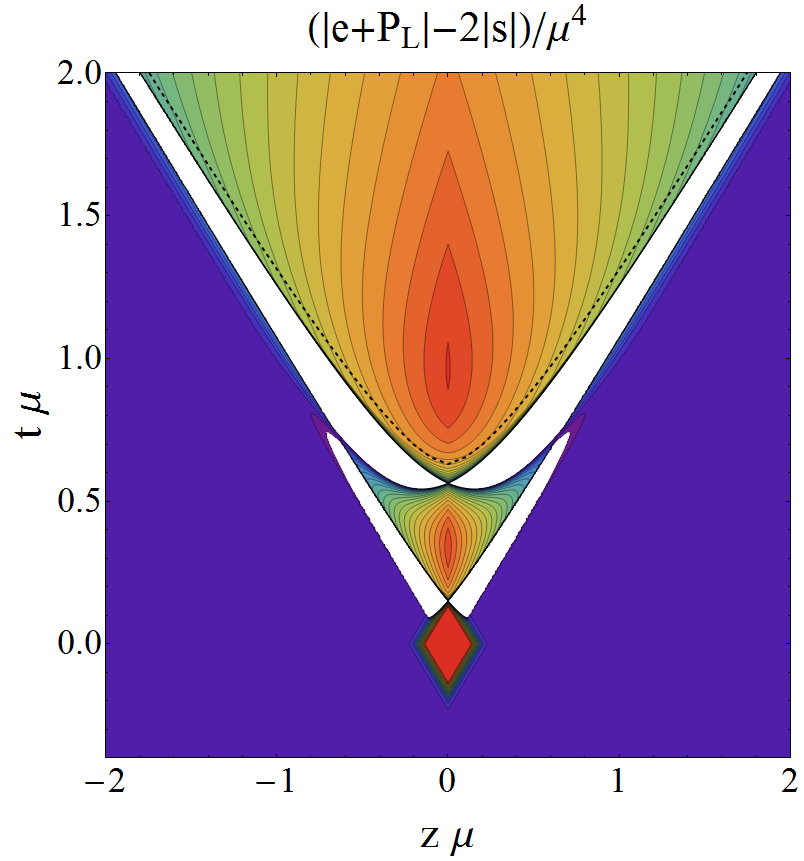} \,\,
  \includegraphics[height=0.52\linewidth]{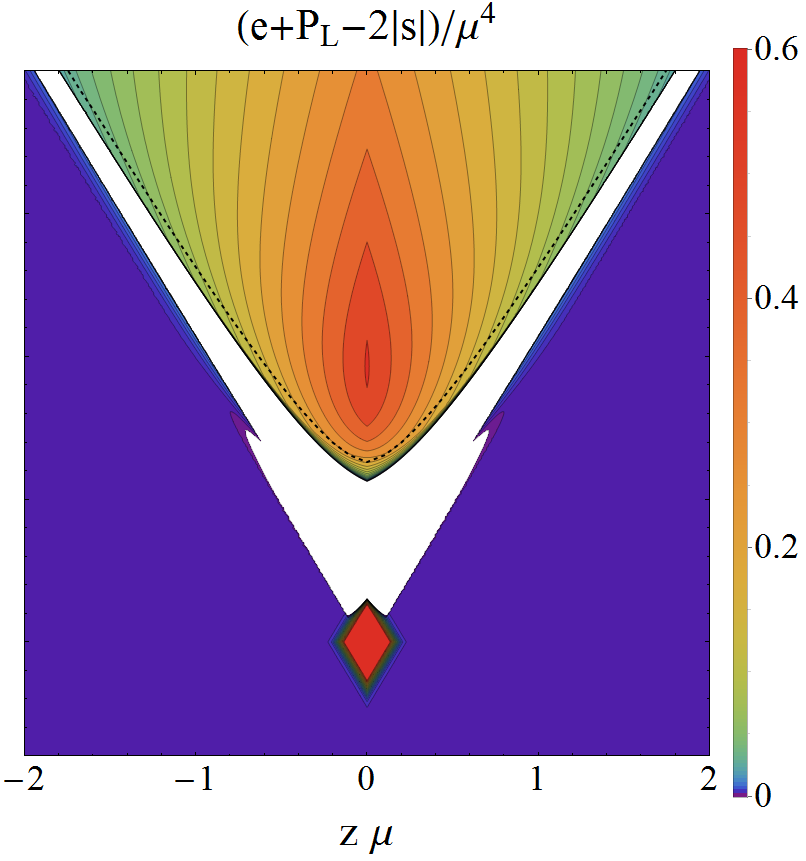}
  \caption{\label{fig:shocks} (Color online) (left) Condition \ref{condition} is plotted for colliding planar shocks, where  $(e,\,P_L,\,s) \equiv \frac{2\pi^2}{N_c^2} (T^{00},\,T^{zz},\,T^{0z})$. A local rest frame does not exist in the white regions, whereas after the black dashed line viscous hydrodynamics provides a good description \cite{Casalderrey-Solana:2013aba}. The region around the origin near $t\mu\approx 0.4$ has $T^{zz}<-T^{00}<0$ in which the rest frame motion is opposite to the direction of the momentum flux $T^{0z}$. (right)  The white regions violate the \ref{NECcondition}. This region is indeed larger than the white regions in the left figure.}
\end{figure*}

Consider the following 2-particle state of the theory of a free
scalar field with mass $m$:
\begin {equation}
   |\Psi\rangle
   \equiv
   \frac{
      |{\rm vac}\rangle
      + \epsilon |\k \k\rangle
      + \epsilon \alpha  |\bzero \bzero\rangle
   }
   {1 + \epsilon^2 + \epsilon^2\alpha^2} ,
\label {eq:Psi}
\end {equation}
where $|\vac\rangle$ is the vacuum state;
$|\k \k\rangle$ is a state of 2 particles each having momentum in the $+z$ direction with
$\k \not= 0$; $|\bzero\bzero\rangle$ is the same state with $\k=0$;
$\epsilon$ is a constant which we shall take to be arbitrarily
small; and $\alpha$ is another constant that we shall fix later.
Calculating the expectation value
$\langle\Psi|{:}T^{\mu\nu}{:}|\Psi\rangle$ of the stress-energy
tensor in this state gives, to first order in $\epsilon$,
\begin {align}
   \langle {:}T^{00}{:} \rangle
   &=
   - \frac{\epsilon\sqrt{2}}{\omega_k V}
     \left[
        \k^2 \cos(2 k\cdot x)
     \right] ,
\\
   \langle {:}T^{zz}{:} \rangle
   &=
   - \frac{\epsilon\sqrt{2}}{\omega_k V}
     \left[
        \omega_k^2 \cos(2 k\cdot x)
        + \alpha m^2 \cos(2 m t)
     \right] ,
\\
   \langle {:}T^{0z}{:} \rangle
   &=
   - \frac{\epsilon\sqrt{2}}{\omega_k V}
     \left[
        \omega_k |\k| \cos(2 k\cdot x)
     \right] ,
\end {align}
where the colons represent normal ordering, which fixes the energy of the vacuum to be zero;
$\omega_k \equiv (\k^2+m^2)^{1/2}$;
and $k\cdot x\equiv -\omega_k t + \k\cdot\x$.
$V$ represents the volume of space.%
\footnote{
   The fact that the state is spread out over all of space is
   inessential.  If one prefers, one could replace each $\k$ in
   $|\k\k\rangle$ by a Gaussian superposition of
   momenta very close to $\k$ (and similarly for $|\bzero\bzero\rangle$)
   in order to localize
   $\langle {:}T^{\mu\nu}{:}\rangle$ in space at each time.
}
The LRF condition is then
\begin {equation}
   \left|
     (\omega_k^2+\k^2) \cos(2 k\cdot x) + \alpha m^2\cos(2 m t)
   \right|
   >
   2 \omega_k |\k| |\cos(2 k\cdot x)| .
\end {equation}
If we now choose the constant $\alpha$ to be
\begin {equation}
   \alpha = - \frac{(\omega_k^2+\k^2)}{m^2} \,,
\end {equation}
for example, then the LRF condition at the origin
$(t,\x)=(0,0)$ becomes simply $0 > 2\omega_k |k_z|$, which
is clearly violated.

Note that the original example in \cite{FullingDavies:1976} had $m=0$ and was constructed to violate the \ref{WECcondition2}. Nevertheless, it only marginally violates \ref{condition} ($|T^{00}+T^{zz}|=2|T^{0z}|$), and is therefore type II in the classification of Hawking and Ellis \cite{Hawking:1973uf}.  Including a non-zero mass makes it possible to have an example of type IV ($|T^{00}+T^{zz}|<2|T^{0z}|$), which furthermore highlights the difference between the LRF and NEC conditions.

There are other regions of spacetime where the LRF condition
is not violated.
The spatial width of regions
of violation is the de Broglie wavelength associated with $\k$.
This is related to the fact that the LRF condition is always
satisfied in the classical limit.

\section{Colliding shocks in $\mathbf{\mathcal{N}=4}$ SYM \label{example}}
\noindent
In the context of gauge/gravity duality, the expectation value of the 
energy-momentum tensor of certain strongly coupled quantum field theories in four spacetime dimensions
(in this case the large-$N_c$ limit of
$\mathcal{N}=4$ SU($N_c$) SYM, with $N_c$ the number of colors) may be calculated from the metric tensor
of classical gravity in five-dimensional asymptotically AdS spacetime.

Of special interest to the study of extremely high-energy nuclear collisions is the question of how field theories equilibrate after the collision of two lumps of matter (`nuclei`), which is relevant to the heavy-ion
program at the Relativistic Heavy Ion Collider (RHIC) and the Large Hadron Collider (LHC). In ${\cal N}=4$ SYM, such collisions are modeled as the collision of two thin planar shock waves \cite{Casalderrey-Solana:2013aba,Chesler:2013lia}, such that any non-trivial dynamics is happening along the $z$ direction, similar to the example described in the previous sections.

\begin{figure*}[t!]
  \centering
  \includegraphics[width=0.45\linewidth]{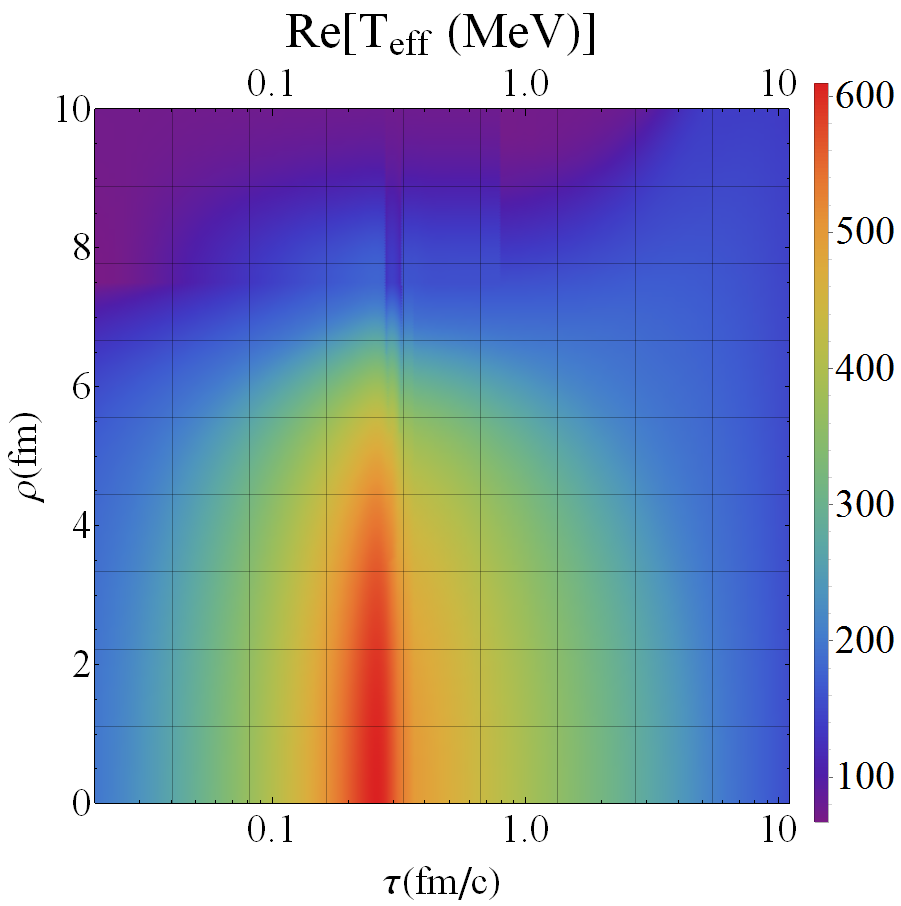}
  \includegraphics[width=0.45\linewidth]{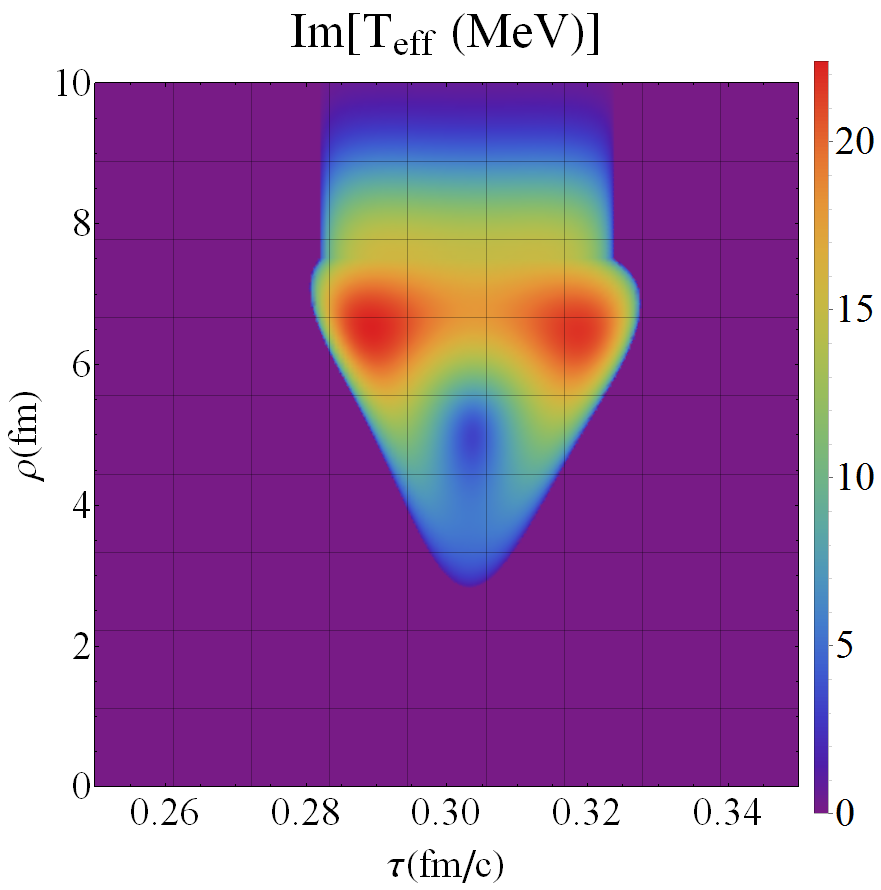}
  \caption{\label{fig:radial} (Color online) We plot the real and imaginary part of the effective temperature: $T_{\text{eff}} = 12 e_{loc}^{1/4}$ of a radially expanding plasma with boost-invariance \cite{vanderSchee:2013pia}. In equilibrium the temperature is naturally real, but we find a region where no local rest frame exists, which is here illustrated by an imaginary part when applying formula \ref{eloc} (note the different axis). Even though no local rest frame exists, the local energy density is continuous, albeit receiving a small imaginary part.}
\end{figure*}

The initial condition for such a collision is given by two single shocks, moving at the speed of light, with an energy-momentum tensor whose only non-zero components are
\begin{equation}
T_{\pm \pm }(z_\pm) = \frac{N_{c}^{2}}{2 \pi^{2}} \, \frac{\mu^3}{\sqrt{2 \pi} w} \, e^{-z_\pm^2/2 w^2},
\end{equation}
where $z$ is the `beam direction', $z_\pm = t\pm z$, $w$ is the width of the sheets, $\mu^3$ the energy per transverse area and the sign depends on the direction of motion of each shock.

When the shocks approach each other the dynamics becomes non-trivial, but the problem can be solved with numerical general relativity using the gravitational dual of this collision. The energy-momentum tensor may be extracted numerically at every spacetime point. A plot of the condition (\ref{condition}) is shown in Figure (\ref{fig:shocks}, left), which demonstrates that (\ref{condition}) is violated in regions close to the light cone. The right of this figure shows an even bigger region where the (\ref{NECcondition}) is violated. While there is a large region where $T^{00}-(T^{0z})^2/T^{zz}\leq 0$, this never coincides with $|T^{0z}|\leq|T^{zz}|$, and hence conditions (\ref{WECcondition2}) and (\ref{NECcondition}) are equivalent in this example.

Importantly, ${\cal N}=4$ SYM is a scale invariant theory and hence all dynamics only depends on dimensionless quantities, such as $\mu w$. Similarly, the size of the region without a rest frame is unlimited, but larger sizes require smaller collision energies. 

In \cite{Casalderrey-Solana:2013sxa} it was shown that the resulting energy-momentum tensor becomes insensitive to the width $w$ if $w\lesssim 0.1/\mu$. The results presented are in this regime, with $\mu w=0.05$, and are hence expected to be a robust outcome for highly Lorentz contracted shocks in strongly coupled quantum theories.

\section{Far from equilibrium matter with radial expansion \label{example2}}
\noindent
One may worry that the violation of the condition (\ref{condition}) studied above is a peculiar feature of the set-up, which in particular is symmetric in the transverse plane. We therefore present a different holographic example, where the expansion is assumed to be boost-invariant in the longitudinal direction and with radial expansion in the transverse plane. The set-up is described in more detail in \cite{vanderSchee:2012qj, vanderSchee:2013pia}.  Here it suffices to say that, as before, we start with perfectly well-defined initial conditions and compute the future stress-energy tensor. This tensor once again has some time of far from equilibrium evolution, after which hydrodynamics applies.

If a local rest frame exists it is interesting to plot the local energy density and velocity of this rest frame. For a boost-invariant stress tensor with radial expansion these can be computed as
\begin{align}
e_{loc}=\frac{1}{2} \left(\sqrt{(T^{\text{rr}}+T^{\tau \tau })^2-4 (T^{\text{$\tau $r}})^2}-T^{\text{rr}}+T^{\tau \tau }\right) \nonumber \\
v_{r,loc}=\frac{T^{rr }+T^{\tau \tau }-\sqrt{(T^{rr }+T^{\tau \tau })^2-4 (T^{\tau r })^2}}{2 T^{\tau r }},
\label{eloc}
\end{align}
where now $\tau$ and $r$ are the usual proper time and radial coordinate, and $v_{r,loc} \equiv u_{r,loc}/u_{\tau,loc}$ the usual rest frame velocity. In this form it is also easy to see when a rest frame cannot be found: if the argument of the square root becomes negative, i.e.\ when  $|T^{\tau\tau}+T^{rr}|<2 |T^{\tau r}|$, conforming to eqn.\ (\ref{condition}).

In figure \ref{fig:radial} we plot the real and imaginary part of the effective temperature,  $T_{\text{eff}} = 12 e_{loc}^{1/4}$. Even though a local energy density is strictly speaking not defined in regions without a local rest frame, we find that eqn.\ (\ref{eloc}) has a smooth continuation when allowing an imaginary part.

We stress that the plots made here are at mid-rapidity and are hence to be compared with a $z=0$ slice in the example of section \ref{example}. There the symmetry guaranteed a rest frame everywhere if $z=0$. Here we relaxed that symmetry and again found regions without a rest frame, indicating that this phenomenon is a general feature of far from equilibrium quantum matter.

\section{Discussion}
\noindent
We would like to mention that there is an elaborate literature on averaged energy conditions \cite{Tipler:1978zz, Klinkhammer:1991ki, Fewster:2002ne, Barcelo:2002bv} and quantum inequalities \cite{Ford:1994bj}, which are believed to hold for all physical states. These averaged conditions integrate a pointwise condition over a complete geodesic, i.e.~the averaged NEC becomes
\begin{equation}
\int_{-\infty}^\infty T_{\mu\nu}K^\mu K^\nu d\lambda > 0,
\end{equation}
where $K^\mu$ is the tangent vector along a null geodesic and $\lambda$ is an affine parameter. The simulations presented in section \ref{example} and \ref{example2} do not allow integration all the way to plus infinity, but integrating to the endpoint of the domain indicates that this condition is easily satisfied. For the colliding shocks this is mainly explained by a large positive contribution on the lightcone, contributing a large amount of positive energy before the region where the NEC is violated is reached.

In the literature a violation of condition (\ref{condition}) has also been observed by a different group, using a completely different setup. Figueras and Wiseman \cite{Figueras:2012rb} report on numerical results for a strongly coupled ${\cal N}$=4 SYM plasma in curved space that flows in response to an adjustable gradient. They find that for small gradients the velocity of matter $v$ is close to the result one would find from fluid dynamics, but for larger gradients the results start to differ until for some critical value of the gradient at which a local rest frame can no longer be defined with subluminal matter velocities.

In conclusion we have shown examples in both free and strongly coupled quantum field theories where regions exist without an LRF. We stress that these regions arose from perfectly well defined initial conditions, whereby violations are the result of a dynamical collision. The examples furthermore highlight differences between the existence of an LRF, the WEC and NEC, where in this case the WEC always implied a rest frame. Lastly, while the collisions studied do not directly correspond to real heavy ion collisions, it is interesting that this analysis suggests there is a serious possibility that regions without a rest frame are briefly created in the RHIC and LHC accelerators.

{{\bf Acknowledgments.}}
We thank Francesco Becattini and Michal P.~Heller for interesting discussions and Frederic Br\"{u}nner for pointing out an error in the original manuscript. This work was supported by the U.S. Department of Energy Office of Science through grant number DE-FG02-03ER41259 and award Nos.\ DE-SC0008132 and DE-SC0007984 as well as a Utrecht University Foundations of Science grant.

\bibliographystyle{apsrev} \bibliography{draftfinalrev}

\begin{thebibliography}{23}
\expandafter\ifx\csname natexlab\endcsname\relax\def\natexlab#1{#1}\fi
\expandafter\ifx\csname bibnamefont\endcsname\relax
  \def\bibnamefont#1{#1}\fi
\expandafter\ifx\csname bibfnamefont\endcsname\relax
  \def\bibfnamefont#1{#1}\fi
\expandafter\ifx\csname citenamefont\endcsname\relax
  \def\citenamefont#1{#1}\fi
\expandafter\ifx\csname url\endcsname\relax
  \def\url#1{\texttt{#1}}\fi
\expandafter\ifx\csname urlprefix\endcsname\relax\def\urlprefix{URL }\fi
\providecommand{\bibinfo}[2]{#2}
\providecommand{\eprint}[2][]{\url{#2}}

\bibitem[{\citenamefont{Hawking and Ellis}(1973)}]{Hawking:1973uf}
\bibinfo{author}{\bibfnamefont{S.}~\bibnamefont{Hawking}} \bibnamefont{and}
  \bibinfo{author}{\bibfnamefont{G.}~\bibnamefont{Ellis}}
  (\bibinfo{year}{1973}).

\bibitem[{\citenamefont{Klinkhammer}(1991)}]{Klinkhammer:1991ki}
\bibinfo{author}{\bibfnamefont{G.}~\bibnamefont{Klinkhammer}},
  \bibinfo{journal}{Phys.Rev.} \textbf{\bibinfo{volume}{D43}},
  \bibinfo{pages}{2542} (\bibinfo{year}{1991}).

\bibitem[{\citenamefont{Roman}(1986)}]{Roman:1986tp}
\bibinfo{author}{\bibfnamefont{T.}~\bibnamefont{Roman}},
  \bibinfo{journal}{Phys.Rev.} \textbf{\bibinfo{volume}{D33}},
  \bibinfo{pages}{3526} (\bibinfo{year}{1986}).

\bibitem[{\citenamefont{Fewster and Roman}(2003)}]{Fewster:2002ne}
\bibinfo{author}{\bibfnamefont{C.~J.} \bibnamefont{Fewster}} \bibnamefont{and}
  \bibinfo{author}{\bibfnamefont{T.~A.} \bibnamefont{Roman}},
  \bibinfo{journal}{Phys.Rev.} \textbf{\bibinfo{volume}{D67}},
  \bibinfo{pages}{044003} (\bibinfo{year}{2003}), \eprint{gr-qc/0209036}.

\bibitem[{\citenamefont{Ford and Roman}(1990)}]{Ford:1990ae}
\bibinfo{author}{\bibfnamefont{L.}~\bibnamefont{Ford}} \bibnamefont{and}
  \bibinfo{author}{\bibfnamefont{T.~A.} \bibnamefont{Roman}},
  \bibinfo{journal}{Phys.Rev.} \textbf{\bibinfo{volume}{D41}},
  \bibinfo{pages}{3662} (\bibinfo{year}{1990}).

\bibitem[{\citenamefont{Dubovsky et~al.}(2006)\citenamefont{Dubovsky, Gregoire,
  Nicolis, and Rattazzi}}]{Dubovsky:2005xd}
\bibinfo{author}{\bibfnamefont{S.}~\bibnamefont{Dubovsky}},
  \bibinfo{author}{\bibfnamefont{T.}~\bibnamefont{Gregoire}},
  \bibinfo{author}{\bibfnamefont{A.}~\bibnamefont{Nicolis}}, \bibnamefont{and}
  \bibinfo{author}{\bibfnamefont{R.}~\bibnamefont{Rattazzi}},
  \bibinfo{journal}{JHEP} \textbf{\bibinfo{volume}{0603}}, \bibinfo{pages}{025}
  (\bibinfo{year}{2006}), \eprint{hep-th/0512260}.

\bibitem[{\citenamefont{Chesler and Yaffe}(2011)}]{Chesler:2010bi}
\bibinfo{author}{\bibfnamefont{P.~M.} \bibnamefont{Chesler}} \bibnamefont{and}
  \bibinfo{author}{\bibfnamefont{L.~G.} \bibnamefont{Yaffe}},
  \bibinfo{journal}{Phys.Rev.Lett.} \textbf{\bibinfo{volume}{106}},
  \bibinfo{pages}{021601} (\bibinfo{year}{2011}), \eprint{1011.3562}.

\bibitem[{\citenamefont{Casalderrey-Solana
  et~al.}(2013)\citenamefont{Casalderrey-Solana, Heller, Mateos, and van~der
  Schee}}]{Casalderrey-Solana:2013aba}
\bibinfo{author}{\bibfnamefont{J.}~\bibnamefont{Casalderrey-Solana}},
  \bibinfo{author}{\bibfnamefont{M.~P.} \bibnamefont{Heller}},
  \bibinfo{author}{\bibfnamefont{D.}~\bibnamefont{Mateos}}, \bibnamefont{and}
  \bibinfo{author}{\bibfnamefont{W.}~\bibnamefont{van~der Schee}},
  \bibinfo{journal}{Phys.Rev.Lett.} \textbf{\bibinfo{volume}{111}},
  \bibinfo{pages}{181601} (\bibinfo{year}{2013}), \eprint{1305.4919}.

\bibitem[{\citenamefont{van~der Schee et~al.}(2013)\citenamefont{van~der Schee,
  Romatschke, and Pratt}}]{vanderSchee:2013pia}
\bibinfo{author}{\bibfnamefont{W.}~\bibnamefont{van~der Schee}},
  \bibinfo{author}{\bibfnamefont{P.}~\bibnamefont{Romatschke}},
  \bibnamefont{and} \bibinfo{author}{\bibfnamefont{S.}~\bibnamefont{Pratt}},
  \bibinfo{journal}{Phys.Rev.Lett.} \textbf{\bibinfo{volume}{111}},
  \bibinfo{pages}{222302} (\bibinfo{year}{2013}), \eprint{1307.2539}.

\bibitem[{\citenamefont{Chesler and Yaffe}(2013)}]{Chesler:2013lia}
\bibinfo{author}{\bibfnamefont{P.~M.} \bibnamefont{Chesler}} \bibnamefont{and}
  \bibinfo{author}{\bibfnamefont{L.~G.} \bibnamefont{Yaffe}}
  (\bibinfo{year}{2013}), \eprint{1309.1439}.

\bibitem[{\citenamefont{Janik and Peschanski}(2006)}]{Janik:2005zt}
\bibinfo{author}{\bibfnamefont{R.~A.} \bibnamefont{Janik}} \bibnamefont{and}
  \bibinfo{author}{\bibfnamefont{R.~B.} \bibnamefont{Peschanski}},
  \bibinfo{journal}{Phys.Rev.} \textbf{\bibinfo{volume}{D73}},
  \bibinfo{pages}{045013} (\bibinfo{year}{2006}), \eprint{hep-th/0512162}.

\bibitem[{\citenamefont{Albacete et~al.}(2008)\citenamefont{Albacete,
  Kovchegov, and Taliotis}}]{Albacete:2008vs}
\bibinfo{author}{\bibfnamefont{J.~L.} \bibnamefont{Albacete}},
  \bibinfo{author}{\bibfnamefont{Y.~V.} \bibnamefont{Kovchegov}},
  \bibnamefont{and} \bibinfo{author}{\bibfnamefont{A.}~\bibnamefont{Taliotis}},
  \bibinfo{journal}{JHEP} \textbf{\bibinfo{volume}{0807}}, \bibinfo{pages}{100}
  (\bibinfo{year}{2008}), \eprint{0805.2927}.

\bibitem[{\citenamefont{Gubser et~al.}(2008)\citenamefont{Gubser, Pufu, and
  Yarom}}]{Gubser:2008pc}
\bibinfo{author}{\bibfnamefont{S.~S.} \bibnamefont{Gubser}},
  \bibinfo{author}{\bibfnamefont{S.~S.} \bibnamefont{Pufu}}, \bibnamefont{and}
  \bibinfo{author}{\bibfnamefont{A.}~\bibnamefont{Yarom}},
  \bibinfo{journal}{Phys.Rev.} \textbf{\bibinfo{volume}{D78}},
  \bibinfo{pages}{066014} (\bibinfo{year}{2008}), \eprint{0805.1551}.

\bibitem[{\citenamefont{Grumiller and Romatschke}(2008)}]{Grumiller:2008va}
\bibinfo{author}{\bibfnamefont{D.}~\bibnamefont{Grumiller}} \bibnamefont{and}
  \bibinfo{author}{\bibfnamefont{P.}~\bibnamefont{Romatschke}},
  \bibinfo{journal}{JHEP} \textbf{\bibinfo{volume}{0808}}, \bibinfo{pages}{027}
  (\bibinfo{year}{2008}), \eprint{0803.3226}.

\bibitem[{\citenamefont{Aref'eva et~al.}(2009)\citenamefont{Aref'eva, Bagrov,
  and Guseva}}]{Aref'eva:2009wz}
\bibinfo{author}{\bibfnamefont{I.~Y.} \bibnamefont{Aref'eva}},
  \bibinfo{author}{\bibfnamefont{A.}~\bibnamefont{Bagrov}}, \bibnamefont{and}
  \bibinfo{author}{\bibfnamefont{E.}~\bibnamefont{Guseva}},
  \bibinfo{journal}{JHEP} \textbf{\bibinfo{volume}{0912}}, \bibinfo{pages}{009}
  (\bibinfo{year}{2009}), \eprint{0905.1087}.

\bibitem[{\citenamefont{Fulling and Davies}(1976)}]{FullingDavies:1976}
\bibinfo{author}{\bibfnamefont{S.~A.} \bibnamefont{Fulling}} \bibnamefont{and}
  \bibinfo{author}{\bibfnamefont{P.~C.~W.} \bibnamefont{Davies}},
  \bibinfo{journal}{Proc. R. Soc. London} \textbf{\bibinfo{volume}{A348}},
  \bibinfo{pages}{393} (\bibinfo{year}{1976}).

\bibitem[{\citenamefont{Davies and Fulling}(1977)}]{Davies:1977yv}
\bibinfo{author}{\bibfnamefont{P.}~\bibnamefont{Davies}} \bibnamefont{and}
  \bibinfo{author}{\bibfnamefont{S.}~\bibnamefont{Fulling}},
  \bibinfo{journal}{Proc.Roy.Soc.Lond.} \textbf{\bibinfo{volume}{A356}},
  \bibinfo{pages}{237} (\bibinfo{year}{1977}).

\bibitem[{\citenamefont{Casalderrey-Solana
  et~al.}(2014)\citenamefont{Casalderrey-Solana, Heller, Mateos, and van~der
  Schee}}]{Casalderrey-Solana:2013sxa}
\bibinfo{author}{\bibfnamefont{J.}~\bibnamefont{Casalderrey-Solana}},
  \bibinfo{author}{\bibfnamefont{M.~P.} \bibnamefont{Heller}},
  \bibinfo{author}{\bibfnamefont{D.}~\bibnamefont{Mateos}}, \bibnamefont{and}
  \bibinfo{author}{\bibfnamefont{W.}~\bibnamefont{van~der Schee}},
  \bibinfo{journal}{Phys.Rev.Lett.} \textbf{\bibinfo{volume}{112}},
  \bibinfo{pages}{221602} (\bibinfo{year}{2014}), \eprint{1312.2956}.

\bibitem[{\citenamefont{van~der Schee}(2013)}]{vanderSchee:2012qj}
\bibinfo{author}{\bibfnamefont{W.}~\bibnamefont{van~der Schee}},
  \bibinfo{journal}{Phys.Rev.} \textbf{\bibinfo{volume}{D87}},
  \bibinfo{pages}{061901} (\bibinfo{year}{2013}), \eprint{1211.2218}.

\bibitem[{\citenamefont{Tipler}(1978)}]{Tipler:1978zz}
\bibinfo{author}{\bibfnamefont{F.~J.} \bibnamefont{Tipler}},
  \bibinfo{journal}{Phys.Rev.} \textbf{\bibinfo{volume}{D17}},
  \bibinfo{pages}{2521} (\bibinfo{year}{1978}).

\bibitem[{\citenamefont{Barcelo and Visser}(2002)}]{Barcelo:2002bv}
\bibinfo{author}{\bibfnamefont{C.}~\bibnamefont{Barcelo}} \bibnamefont{and}
  \bibinfo{author}{\bibfnamefont{M.}~\bibnamefont{Visser}},
  \bibinfo{journal}{Int.J.Mod.Phys.} \textbf{\bibinfo{volume}{D11}},
  \bibinfo{pages}{1553} (\bibinfo{year}{2002}), \eprint{gr-qc/0205066}.

\bibitem[{\citenamefont{Ford and Roman}(1995)}]{Ford:1994bj}
\bibinfo{author}{\bibfnamefont{L.}~\bibnamefont{Ford}} \bibnamefont{and}
  \bibinfo{author}{\bibfnamefont{T.~A.} \bibnamefont{Roman}},
  \bibinfo{journal}{Phys.Rev.} \textbf{\bibinfo{volume}{D51}},
  \bibinfo{pages}{4277} (\bibinfo{year}{1995}), \eprint{gr-qc/9410043}.

\bibitem[{\citenamefont{Figueras and Wiseman}(2013)}]{Figueras:2012rb}
\bibinfo{author}{\bibfnamefont{P.}~\bibnamefont{Figueras}} \bibnamefont{and}
  \bibinfo{author}{\bibfnamefont{T.}~\bibnamefont{Wiseman}},
  \bibinfo{journal}{Phys.Rev.Lett.} \textbf{\bibinfo{volume}{110}},
  \bibinfo{pages}{171602} (\bibinfo{year}{2013}), \eprint{1212.4498}.

\end{thebibliography}

\end{document}